\documentclass[aps,prc,showpacs,superscriptaddress,groupedaddress]{revtex4-1}
\usepackage{amsmath,amssymb,amsbsy,bm}
\usepackage{graphicx}
\usepackage{comment}
\usepackage{float}
\usepackage[colorlinks=true,linkcolor=blue,citecolor=blue,urlcolor=blue]{hyperref}

\begin{document}

\title{Calculating $n$-Point Charge Correlations in Evolving Systems}
\author{Scott Pratt}
\affiliation{Department of Physics and Astronomy and National Superconducting Cyclotron Laboratory\\
Michigan State University, East Lansing, MI 48824~~USA}
\date{\today}


\begin{abstract}
In dynamic systems, charge susceptibilities and local charge correlations change with time. These changes are accompanied by non-local correlations which spread diffusively with time and are constrained by local charge conservation. Assuming the local features of the correlation, which for a gas would be the correlation of charges within the same particle, are equilibrated, a diagrammatic formalism is presented for calculating the evolution of the associated non-local correlations. These provide correlations of $n$ density operators at different positions for arbitrary $n$. The techniques were developed with an eye towards relativistic heavy-ion collisions, and can account for correlations indexed by up, down and strange charges. Understanding the evolution of such correlations is crucial if one is to interpret measurements of charge fluctuations from the Relativistic Heavy-Ion Collider.  
\end{abstract}

\maketitle
\section{Introduction}

Correlations and fluctuations of conserved charges play a central role in heavy-ion collisions. Charge fluctuations represent defining property of any bulk system, particularly in characterizing phase transitions. As temperatures rise above $\approx 160$ MeV, matter undergoes a transition from a hadronic gas to a strongly interacting plasma of quarks and gluons, the quark-gluon plasma (QGP). For neutral matter, equal numbers of particles and anti-particles, lattice gauge theory has shown that the transition is a smooth crossover occurring within a rather narrow window of temperatures, $150\lesssim T\lesssim 180$ MeV. For temperatures below 150 MeV charge susceptibilities from lattice calculations are consistent with expectations for a weakly interacting gas of hadrons, and for temperatures above 180 MeV, they become consistent with a weakly interacting gas of light up, down and strange quarks \cite{Koch:2008ia,Bzdak:2012an,Borsanyi:2011sw,Bazavov:2012jq,Bazavov:2014xya}. This consistency is rather surprising given the experimental evidence that the system behaves like a nearly ideal liquid, with mean free paths on the order of the thermal wavelength \cite{Romatschke:2007mq,Song:2010mg,Pratt:2015zsa,Auvinen:2017fjw}. Fluctuations between baryon charge and strangeness \cite{Koch:2005vg}, and baryon fluctuations of third or fourth order, e.g. $\langle \delta Q^3\rangle$ or $\langle\delta Q^4\rangle$, are especially illuminating \cite{Athanasiou:2010kw}. They suggest that charges fluctuate in units of one third baryon number rather than in units of baryon number once temperatures rise above the aforementioned window \cite{Bazavov:2014xya}. At finite baryon density, where lattice calculations struggle due to a sign problem, the possibility exists for a first-order phase transition, which would culminate at a critical point \cite{Stephanov:2008qz,Stephanov:2011pb,Vovchenko:2015pya}. The baryon density of the critical point might be several times normal nuclear density and the critical temperature would likely be moderately less than the temperature window quoted above for the smooth transition at zero baryon density. Baryon fluctuations should represent identifying properties of such a phase transition, particularly near the critical point. Fluctuations of baryon number in high energy heavy ion collisions have been analyzed as a function of beam energy at the Relativistic Heavy Ion Collider (RHIC) \cite{Aggarwal:2010wy,Adamczyk:2013dal,Adamczyk:2014fia,Adamczyk:2017wsl,McDonald:2013aoa}. At the highest RHIC energies, the incoming beams are insufficiently stopped to contribute large numbers of baryons to the mid-rapidity region. Combined with rampant particle production at such energies, net baryon densities are much lower than entropy densities and experiments are able to investigate the properties of matter with nearly zero baryon chemical potential, $\mu_B\approx 0$. Baryon densities increase for lower beam energies, which provides the opportunity to study the properties of matter as a function of baryon density at high temperature. 

Unfortunately, interpreting charge fluctuations is greatly complicated by the dynamic nature of the collision. A system's baryon density and temperature traverse a swath through the density-temperature plane. Thus, any measurement reflects on the bulk properties over a range of density and temperature. Further, the short lifetime of these environments restricts charge fluctuations within any volume from attaining equilibrium values because charge is locally conserved and requires significant time to diffuse across the volume. Any quantitative model of charge fluctuations must therefore describe the evolution of charge correlations, $\langle \delta\rho(\bm{r}_1,t)\cdots\delta\rho(\bm{r}_n,t)\rangle$, in order to understand how the equilibrium properties one would study in a static system would become manifest in the finite-size and finite-volume systems created in heavy-ion collisions, where measurement of the conserved charges are confined to the final state.

Models have addressed the challenges outlined in the previous paragraph, but mainly for two-point correlations \cite{Stephanov:2017ghc,Stephanov:2017wlw,Pratt:2017oyf,Pratt:2018ebf}. As reviewed in the next section, two-point correlations can be split into two pieces. The first piece is the short-range contribution. This is when the two density operators in the correlation, $\langle\delta\rho(\bm{r}_1,t)\delta\rho(\bm{r}_2,t)\rangle$, refer to charges within the same particle. Throughout this paper, the word ``particle'' can be extended to any short-range feature that might locally equilibrate. For example, in a hadron gas this would be the case where the two density operators referred to the same hadron, and for a QGP gas, this would be when the two density operators address the same quark. Assuming knowledge of the local part of the correlation part, e.g. assuming a chemically equilibrated gas, the remainder of the correlation function is constrained by the fact that the net correlation function must integrate to zero due to local charge conservation. If one assumes that charges move diffusively, the evolution of the non-local part of the correlation function can be modeled by the diffusion equation with a source term given by the rate of change of the local part. Such an approach was superimposed onto a hydrodynamic description of a heavy-ion collision in \cite{Pratt:2017oyf} and then extended to include a hybrid hydrodynamic model interfaced to a hadronic simulation that simulated the break-up stage of a heavy-ion collision \cite{Pratt:2018ebf,Pratt:2019pnd}. These approaches roughly reproduced several experimental measures of charge correlations from the STAR collaboration at RHIC, which should translate into reproducing experimental measures of charge fluctuations to order $\langle\delta Q^2\rangle$, because fluctuations are determined by integrating over the correlations. 

The aim of this work is to provide a theoretical foundation to extend the treatment of two-point correlations to three-point, four-point and $n$-point correlations. This is more difficult than the case for two-point correlations. In that case the correlation was divided into a local part, where the two density operators referred to the same particle, and a non-local part. For three-point correlations, $\langle\delta\rho_a(\bm{r}_1,t)\delta\rho_b(\bm{r}_2,t)\delta\rho(\bm{r}_3,t)\rangle$, one must consider three cases: where all three density operators refer to the same particle, where two of the three refer to the same particle, and where all three refer to different particles. Even if one makes an assumption about the local part for the three-point function, i.e. it reflects chemical equilibrium, one must understand how to split the remaining correlation over the other two possibilities. Four-point and higher correlations offer even more possible splittings. Section \ref{sec:threepoint} shows how three-point correlations can be modeled. The question of how to spread correlations amongst the various splittings is answered by assuming that a differential charge results in differential changes in the various species according to equilibrium. This then allows correlations of order $\delta\rho^3$ due to a two-point function to be determined by the two-point function contribution to correlations of order $\delta\rho^2$.  The algebra in Sec. \ref{sec:threepoint} is rather lengthy, and a similar exposition for $n$-point correlations with $n>3$ would be much more so. Fortunately, the expressions can be represented diagrammatically. A diagrammatic description, which is extendible to higher $n$, is presented in Sec. \ref{sec:npoint}. 

In \cite{Pratt:2017oyf} the diffusion equation for the two-point correlation function was addressed by noting the equivalence with a random walk. In Sec. \ref{sec:algorithms} the benefits of a random-walk algorithm vs. a mesh-based description of the correlation function is discussed. In a quark gas, up, down and strange quarks represent good quasi-particles and the three-by-three diffusivity tensor is diagonal. In a hadron gas, hadrons carry multiple quarks and the diffusivity tensor, just like the susceptibility, is no longer diagonal in the $u,d,s$ basis. A strategy for applying a random walk algorithm in a situation where the diffusivity tensor is not diagonal is also provided in Sec. \ref{sec:algorithms}.

The final section, Sec. \ref{sec:applicability}, presents a discussion of the applicability of the relations from Sec.s \ref{sec:threepoint} and \ref{sec:npoint}. The role of assuming chemical equilibrium is emphasized. Finally, strategies are presented for handling both local and non-local contributions to the susceptibility. Critical phenomena involves correlation on longer length scales and seems well suited for hydrodynamic treatments \cite{Stephanov:2017ghc,Stephanov:2017wlw,Stephanov:2012ki,Pratt:2017lce,Paech:2003fe,Paech:2005cx,Nahrgang:2011mg,Kapusta:2014dja,Kapusta:2012sd,Young:2014pka,Ling:2013ksb}. Phase separation dynamics might also be addressed with such an approach \cite{Steinheimer:2013xxa,Steinheimer:2013gla,Steinheimer:2012gc,Randrup:2010ax,Heiselberg:AnnPhys,Heiselberg:1988oha,Napolitani:2014ima,Chomaz:2003dz,Borderie:2001jg,Colonna:2002ti,Guarnera:1996svb}. However, hydrodynamics, noisy or not, is a clumsy means by which to model the correlation of a particle with itself because hydrodynamics is based on gradients, which implies that correlations have a length scale greater than the inter-particle separation. Section \ref{sec:applicability} describes the possibility of combining a hydrodynamics approach to account for the non-local contribution to the susceptibility and the formalism presented here to account for the local part. This study considers only evolving the correlations in coordinate space, whereas measurements are restricted to the asymptotic momenta. Techniques for translating correlations to momentum space has been described and implemented in \cite{Pratt:2011bc,Pratt:2015jsa}, and it would be straight-forward to extend these methods to project $n-$point correlations into momentum space. Implementations of the formalism presented here will be deferred for another study.

\section{Two-point functions}
\label{sec:twopoint}
Before launching into a formalism for three- and four-point functions, that for two-point functions is reviewed here. This has been appeared in \cite{Pratt:2017oyf} and applied to a hydrodynamic evolution of a heavy-ion collision in \cite{Pratt:2017oyf,Pratt:2018ebf,Pratt:2019pnd}.

First, the definitions,
\begin{eqnarray}
\label{eq:C2def}
\mathcal{C}^{\rm(tot)}_{ab}({\bm r}_1,{\bm r}_2,t)&=&\langle \delta\rho_a({\bm r}_1,t)\delta\rho_b({\bm r}_2,t)\rangle\\
\nonumber
&=&\chi^{(2)}_{ab}(\bm{r}_{12},t)\delta({\bm r}_1-\bm{r}_2)+C^{(1;1)}_{a;b}(\bm{r}_1,\bm{r}_2,t),\\
\nonumber
\bm{r}_{ij}&\equiv&(\bm{r}_i+\bm{r}_j)/2.
\end{eqnarray}
The subscript $a$ denotes the various charges, perhaps the $u,d,s$ charges on quarks. Here, $\delta\rho_a=\rho_a-\langle\delta\rho_a\rangle$ so that $\langle\delta\rho_a\rangle=0$. The subscripts on the correlations indicate whether the charges are on the same particle, as $\chi^{(2)}(\bm{r},t)$ describes the contributions where both charges come from the same point, or from the same particle, whereas $C^{(1;1)}$ encapsulates the contribution to the correlation when the charges are on different particles.  The semicolon in $C^{(1;1)}_{a;b}$ emphasizes that the two charges $a$ and $b$ are not on the same particle. For the considerations of this paper, it will be assumed that the local part is understood, i.e. it could be the equilibrated susceptibility if the particles are well defined and are in chemical equilibrium. In the gaseous limit, $\chi^{(2)}_{ab}$ is the correlation of the charges within a particle,
\begin{eqnarray}
\chi^{(2)}_{ab}&=&\sum_s n_s q_{s,a}q_{s,b},
\end{eqnarray}
where $n_s$ is the number density of species $s$, and $q_{s,a}$ is the charge of type $a$ on a particle of species $s$. For the example of a hadron gas, the contribution from $\pi^+$ mesons to $\chi^{(2)}_{ud}$ is $-n_{\pi+}$, where $n_{\pi+}$ is the density of $\pi^+$ mesons.  The negative sign ensues because the $\pi^+$ meson has an up quark and an anti-down quark. Even for individual quarks, one finds a contribution to $\chi^{(2)}$ from the correlations of quarks with themselves. For a gas of quarks, $\chi^{(2)}_{ss}=n_{s}+n_{\bar{s}}$, the density of strange plus that of anti-strange quarks. In this paper, the quasi-particles that carry charge will be referred to as particles. Particles could refer to point charges, hadrons, atoms, molecules, or could even include a local polarization cloud. The non-local part, $C^{(1;1)}$, will diffuse and spread over large relative coordinates. Providing the theoretical structure for calculating  the evolution of $C^{(1;1)}$, for the case of two-particle correlations, and $C^{(1;1;1)}$ or $C^{(1;1;1;1)}$ for three- or four-particle correlations, is the principal goal of this paper. The local correlation, whose strength is $\chi^{(n)}$, will be assumed to be given, by assuming local chemical equilibrium.

The evolution of the correlation is guided by the equation,
\begin{eqnarray}
\label{eq:D1C}
D_{1}C^{\rm(tot)}_{ab}(\bm{r}_1,t_1,\bm{r}_2,t_2)&=&-\langle[\nabla_1\cdot{\bm j}_a(\bm{r}_1,t_1)]\delta\rho_b(\bm{r}_2,t_2)\rangle \\
\nonumber
D_i&\equiv&\frac{\partial}{\partial t_i}+{\bm v}(\bm{r}_i,t_i)\cdot\nabla_i+\nabla_i\cdot{\bm v}({\bm r}_i,t_i).
\end{eqnarray}
Here, $\bm{v}$ is the local velocity of the fluid, and $\bm{j}_a$ is the current measured in the fluid frame, i.e. it neglects the part of the current from $\delta\rho_a\bm{v}$. The definition of $D_i$ differs from the usual definition of a co-moving derivative because of the presence of the term $\nabla\cdot\bm{v}$. That term accounts for the current $\bm{j}$ being measured relative to the local frame of the fluid. If one were to include the term $\delta\rho\bm{v}$ to the current, this additional contribution to $D_i$ would not be necessary. With this definition, $\bm{j}$ can be considered as the diffusive contribution to the current, i.e. it ignores the part from simple fluid movement.  Because the r.h.s of Eq. (\ref{eq:D1C}) is a divergence, this represents local charge conservation. A corresponding equation is also true for $D_2$. To propagate the equal-time correlation forward,
\begin{eqnarray}
C^{(1;1)}_{a;b}(\bm{r}_1+\bm{v}_1dt,t_1+dt,\bm{r}_2+\bm{v}_2dt,t_2+dt)&=&
C^{(1;1)}_{a;b}(\bm{r}_1,t_1,\bm{r}_2,t_2)+dt D_tC^{(1;1)}_{a;b}(\bm{r}_1,t_1,\bm{r}_2,t_2),\\
\nonumber
D_t&=&D_1+D_2,\\
\nonumber
D_t C^{\rm(tot)}_{ab}({\bm r}_1,{\bm r}_2,t)&=& \delta(\bm{r}_1-\bm{r}_2)D_t\chi^{(2)}_{ab}(\bm{r}_{12},t)
+D_t C^{(1;1)}_{ab}(\bm{r}_1,\bm{r}_2,t),\\
\nonumber
\bm{r}_{12}&\equiv&(\bm{r}_1+\bm{r}_2)/2,
\end{eqnarray}
or in terms of $C^{(1;1)}$,
\begin{eqnarray}
\label{eq:evolveC2}
D_t C^{(1;1)}_{a;b}(\bm{r}_1,\bm{r}_2,t)&=&-\langle[\nabla_1\cdot\bm{j}_a(\bm{r}_1,t)]\delta\rho_b(\bm{r}_2,t)\rangle
-\langle\delta\rho_a(\bm{r}_1,t)\nabla_2\cdot\bm{j}_b(\bm{r}_2,t)\rangle\\
\nonumber
&&+S^{(2)}_{ab}(\bm{r}_{12},t)\delta(\bm{r}_1-\bm{r}_2),\\
\nonumber
S^{(2)}_{ab}(\bm{r}_{12},t)&=&-D_t\chi^{(2)}_{ab}(\bm{r}_{12},t).
\end{eqnarray}
The last term, with $S^{(2)}_{ab}(\bm{r}_{12},t)$, behaves like a source function for $C^{(1;1)}$,
\begin{eqnarray}
\int d^3r_1~d^3r_2~C^{(1;1)}_{a;b}(\bm{r}_1,\bm{r}_2,t)&=&\int_{-\infty}^t dt'd^3r'~ S^{(2)}_{ab}(\bm{r}_{12},t').
\end{eqnarray}
The source term contributes to the strength of the correlation at $\bm{r}_1-\bm{r}_2=0$. For a small fluid element of volume $\delta V$ that expands with the fluid, the source contributes when the product $\chi\delta V$ changes with time. This is a consequence of the definition of $D_t$ including the $\nabla\cdot\bm{v}$ term, because the usual comoving derivative, $\partial_t+\bm{v}\cdot\nabla$, acting on $\delta V$ gives
\begin{eqnarray}
\left[\partial_t+\bm{v}\cdot\nabla\right]\delta V&=&(\nabla\cdot\bm{v})\delta V.
\end{eqnarray}
For ideal hydrodynamics, the entropy within $\delta V$, which equals $s\delta V$, would remain constant. In that case,
\begin{eqnarray}
\label{eq:chiovers}
D_t \chi^{(2)}_{ab}(\bm{r},t)&=&s\left[\partial_t+\bm{v}\cdot\nabla\right]\left(\frac{\chi_{ab}^{(2)}(\bm{r},t)}{s(\bm{r},t)}\right),
\end{eqnarray}
and one can see that the source term is principally a function of whether the ratio $\chi/s$ rises or falls as one moves with the fluid. If entropy is not conserved, the source term differs somewhat. In the treatments of \cite{Pratt:2011bc,Pratt:2019pnd,Pratt:2011bc} the hydrodynamic evolution was viscous, and the source term was calculated with the full expression given in Eq. (\ref{eq:evolveC2}). Nonetheless, the approximate form in Eq. (\ref{eq:chiovers}) is insightful, as plotting $\chi_{ab}/s$ as a function of temperature describes at what points in the trajectory the source term becomes significant. Further, this ratio can be calculated in lattice gauge theory \cite{Pratt:2015jsa}.

Although the expressions involve two powers of the density, the evolution of $C^{(1;1)}$ is described by a linear equation including a source term. For each differential contribution to the source function, $-D_t\chi^{(2)}_{ab} d^3r~dt$, one can solve for its contribution of $C^{1,1}$. Finally, one can sum each contribution by integrating over the source function. If the evolution is diffusive, $\bm{j}_a(\bm{r})=-\mathcal{D}_{ab}(\bm{r})\nabla \delta\rho_b(\bm{r})$, where $\mathcal{D}$ is the diffusivity tensor. If the diffusivity tensor is diagonal each of these contributions can be represented by two sample charges $a$ and $b$ undergoing a random walk with the parameters of the random walk set by the diffusivity tensor. The positions of the two charges can then be used to construct the correlation function in coordinate space. It is then straight-forward to design a Monte Carlo procedure to generate pairs for each contribution from the source function at some point $\bm{r}_s$ and time $t_s$. Because only the charges originating from the same source point are correlated with one another, there is no combinatoric noise to overcome. This approach was applied in \cite{Pratt:2017oyf,Pratt:2018ebf,Pratt:2019pnd}. A method for handling non-diagonal diffusivity tensors is provided in Sec. \ref{sec:algorithms}.

\section{Three-Point Correlators}
\label{sec:threepoint}

Without loss of generality, the three-point correlator can be written as
\begin{eqnarray}
\label{eq:3pointparts}
\mathcal{C}^{\rm(tot)}_{abc}(\bm{r}_1,\bm{r}_2,\bm{r}_3,t)&=&\langle\delta\rho_a(\bm{r}_1,t)\delta\rho_b(\bm{r}_2,t)\delta\rho(\bm{r}_3,t)\rangle\\
\nonumber
&=&C^{(1;1;1)}_{a;b;c}(\bm{r}_1,\bm{r}_2,\bm{r}_3,t)+C^{(2;1)}_{ab;c}(\bm{r}_{12},\bm{r_3},t)\delta(\bm{r}_1-\bm{r}_2)\\
\nonumber
&&+C^{(2;1)}_{ac;b}(\bm{r}_{13},\bm{r_2})\delta(\bm{r}_1-\bm{r}_3,t)
+C^{(2;1)}_{bc;a}(\bm{r}_{23},\bm{r_1})\delta(\bm{r}_2-\bm{r}_3,t)\\
\nonumber
&&+\chi^{(3)}_{abc}(\bm{r}_{123},t)\delta(\bm{r}_{12}-\bm{r}_3)\delta(\bm{r}_1-\bm{r}_2).
\end{eqnarray}
Here, $\bm{r}_{123}\equiv(\bm{r}_1+\bm{r}_2+\bm{r}_3)/3$. The correlation $C^{(1;1;1)}$ describes correlations when all three positions are different, i.e. the density operators refer to different particles, and $C^{(2;1)}_{ab;c}$ describes the correlations when two positions are the same, i.e. the two charges $a$ and $b$ are on the same particle and $c$ is on a separate particle. The correlation when all three points are the same, or all three charges are on the same particle, is described by $\chi^{(3)}_{abc}$. Just as with $\chi^{(2)}_{ab}$, this will be identified as the equilibrium susceptibility here. If the particles are molecules, assigning $\chi^{(n)}$ as the equilibrium susceptibility represents an assumption of chemical equilibrium.

The correlator, $C^{(2;1)}_{ab;c}(\bm{r}_{12},\bm{r}_3,t)$, describes the correlation between a charge of type $c$ at $\bm{r}_3$ and a particle at position $\bm{r}_{12}$ carrying a product of charges $Q_aQ_b$. Here, we show that it is directly determined by $C^{(1;1)}_{d;c}(\bm{r}_{12},\bm{r}_3,t)$ and the susceptibilities. To demonstrate this relation, one can consider a particle of species $s$. The charge $\delta Q_d$ due to the increased probability of having a particle $\delta N_s$, is
\begin{eqnarray}
\label{eq:derivNofQ1}
\delta Q_d&=&\sum_s\delta N_s q_{s,d},
\end{eqnarray}
where $q_{sd}$ is the charge of type $d$ on the particle of type $s$. If the particle probability is equilibrated in response to the small charge,
\begin{eqnarray}
\label{eq:derivNofQ2}
\delta N_s&=&\langle N_s\rangle \delta\mu_a q_{s,a},
\end{eqnarray}
where $\delta \mu_a$ is the chemical potential inspired by the small charges, divided by the temperature. One can insert Eq. (\ref{eq:derivNofQ2}) into (\ref{eq:derivNofQ1}),
\begin{eqnarray}
\delta Q_d&=&\sum_s q_{s,d} \langle N_s\rangle \delta\mu_a q_{s,a}\\
\nonumber
&=&V\chi^{(2)}_{da}\delta \mu_a,\\
\nonumber
\delta\mu_a&=&\frac{1}{V}[\chi^{(2)}]^{-1}_{ab}\delta Q_b,
\end{eqnarray}
where $[\chi^{(2)}]^{-1}_{ab}$ is the inverse two-point susceptibility matrix. Inserting this into Eq. (\ref{eq:derivNofQ2}),
\begin{eqnarray}
\label{eq:dNh}
\delta N_s&=&\langle n_s\rangle q_{s,a}[\chi^{(2)}]^{-1}_{ab}\delta Q_b.
\end{eqnarray}
This expresses how many extra particles of type $s$, $\delta N_s$, one would generate in a volume when a small charge, $\delta Q_a$, is added to the volume.

One can now calculate the additional product of charges $\delta(Q_aQ_b)$ due to $\delta Q_c$. To that end, one can consider a small volume $\delta V$ restricting the position $\bm{r}_{12}$. The delta function, should not be of zero extent, but should have a range large enough to fit in a quasi-particle. 
\begin{eqnarray}
\int_{\in \delta V} d^3r_{12}~d^3(\bm{r}_1-\bm{r}_2)~\delta\rho_a(\bm{r}_1)\delta\rho_b(\bm{r}_2) \delta(\bm{r}_1-\bm{r}_2)&=&\frac{1}{\delta V}\delta (Q_aQ_b)\\
\nonumber
&=& \frac{1}{\delta V} \sum_s q_{s,a}q_{s,b}\delta N_s\\
\nonumber
&=&\frac{1}{\delta V}\sum_s\langle n_s\rangle q_{s,a}q_{s,b}q_{s,d}[\chi^{(2)}]^{-1}_{dc}\delta Q_c\\
\nonumber
&=&\chi^{(3)}_{abd}(\bm{r}_{12},t)[\chi^{(2)}]^{-1}_{dc}(\bm{r}_{12},t)\delta\rho_c(\bm{r}_{12},t).
\end{eqnarray}
Here, $\delta(Q_aQ_b)$ refers to the charges inside the volume $\delta V$. It is indeed this product of charges in a single particle at $\bm{r}_{12}$ that is described by $C^{(2;1)}_{ab;c}(\bm{r}_{12},\bm{r}_3,t)$. Thus,
\begin{eqnarray}
\label{eq:Ldef}
C^{(2;1)}_{ab;c}(\bm{r}_{12},\bm{r}_3,t)&=&L^{(2)}_{ab,e}(\bm{r}_{12},t) C^{(1;1)}_{e;c}(\bm{r}_{12},\bm{r}_3,t),\\
\nonumber
L^{(2)}_{ab,e}(\bm{r}_{12},t)&\equiv&\chi^{(3)}_{abd}(\bm{r}_{12},t)[\chi^{(2)}]^{-1}_{de}(\bm{r}_{12},t).
\end{eqnarray}
By assuming that $\chi^{(3)}$ is consistent with chemical equilibrium, all mention of the individual particles and their charges has disappeared, and $C^{(2;1)}_{ab;c}(\bm{r}_{12},\bm{r}_3,t)$ is determined by the correlation and the susceptibilities evaluated at $\bm{r}_{12}$. For future reference, one can readily show that for any product of $m$ charge densities $\rho_a\cdots\rho_c$,
\begin{eqnarray}
\label{eq:Ldefgeneral}
C^{(m;\cdots)}_{a\cdots c;\cdots}(\bm{r},\cdots,t)&=&L^{(m)}_{a\cdots c,d}((\bm{r},t)C^{(1;\cdots)}_{d;\cdots}(\bm{r},\cdots,t),\\
\nonumber
L^{(m)}_{a\cdots c,d}(\bm{r},t)&=&\chi^{(m+1)}_{a\cdots c,e}(\bm{r},t)[\chi^{(2)}(\bm{r},t)]^{-1}_{ed}.
\end{eqnarray}

Our principal goal is to determine the evolution of $C^{(1;1;1)}_{a;b;c}(\bm{r}_1,\bm{r}_2,\bm{r}_3,t)$. Assuming that the two-point correlation $C^{(1;1)}_{a;b}$ was already determined using the methods of Sec. \ref{sec:twopoint}, all terms from the r.h.s. of Eq. (\ref{eq:3pointparts}) involving two-point functions can be calculated from Eq. (\ref{eq:Ldef}). Here, we first solve for $D_t=D_1+D_2+D_3$ of the l.h.s. of the equation, i.e. $D_t$ acting on the total correlation. Then applying $D_t$ to the r.h.s. will provide an expression for $D_tC^{(1;1;1)}$. 

Before applying $D_t$ to the l.h.s. of Eq. (\ref{eq:3pointparts}), one can surround the points $\bm{r}_1$, $\bm{r}_2$ and $\bm{r}_3$ with surfaces and consider the net correlation of the product of charges within the enclosing volumes $V_1$ $V_2$ and $V_3$,
\begin{eqnarray}
\langle \delta Q_a\delta Q_b\delta Q_c\rangle
&=&\int_{V_1V_2V_3} d^3r_1~d^3r_2~d^3r_3~\langle\delta\rho_a(\bm{r}_1,t)\delta\rho_b(\bm{r}_2,t)\delta\rho_c(\bm{r}_3,t)\rangle
\end{eqnarray}
The rate of change of $C_{abc}$ is determined by the rate at which charge flows out of the small encircling volumes,
\begin{eqnarray}
\label{eq:QQQlong}
\frac{d}{dt}\langle \delta Q_a\delta Q_b\delta Q_c\rangle_V&=&
-\int d^3r_2d^3r_3d\bm{A}_1\cdot\langle\bm{j}_a(\bm{r}_1,t)\delta\rho_b(\bm{r}_2,t)\delta\rho_c(\bm{r}_3,t)\rangle'\\
\nonumber
&&-\int d^3r_1d^3r_3d\bm{A}_2\cdot\langle\bm{j}_b(\bm{r}_2,t)\delta\rho_a(\bm{r}_1,t)\delta\rho_c(\bm{r}_3,t)\rangle'\\
\nonumber
&&-\int d^3r_2d^3r_3d\bm{A}_3\cdot\langle\bm{j}_c(\bm{r}_3,t)\delta\rho_a(\bm{r}_1,t)\delta\rho_b(\bm{r}_2,t)\rangle'\\
\nonumber
&&-\int d^3r_3d\bm{A}_{12}L^{(2)}_{ab,d}(\bm{r}_{12})\cdot\langle\bm{j}_d(\bm{r}_{12},t)\delta\rho_c(\bm{r}_3,t)\rangle'\\
\nonumber
&&-\int d^3r_2d\bm{A}_{13}L^{(2)}_{ac,d}(\bm{r}_{13})\cdot\langle\bm{j}_d(\bm{r}_{13},t)\delta\rho_b(\bm{r}_2,t)\rangle'\\
\nonumber
&&-\int d^3r_1d\bm{A}_{23}L^{(2)}_{bc,d}(\bm{r}_{23})\cdot\langle\bm{j}_d(\bm{r}_{23},t)\delta\rho_a(\bm{r}_1,t)\rangle'\\
\nonumber
&&-\int d^3r_{23}d\bm{A}_{1}L^{(2)}_{bc,d}(\bm{r}_{23})\cdot\langle\delta\rho_d(\bm{r}_{23},t)\bm{j}_a(\bm{r}_1,t)\rangle'\\
\nonumber
&&-\int d^3r_{13}d\bm{A}_{2}L^{(2)}_{ac,d}(\bm{r}_{13})\cdot\langle\delta\rho_d(\bm{r}_{13},t)\bm{j}_b(\bm{r}_2,t)\rangle'\\
\nonumber
&&-\int d^3r_{12}d\bm{A}_{3}L^{(2)}_{ab,d}(\bm{r}_{12})\cdot\langle\delta\rho_d(\bm{r}_{12},t)\bm{j}_c(\bm{r}_3,t)\rangle'.
\end{eqnarray}
Here, the prime on the averages $\langle\cdots\rangle'$ restricts the integrals to not include charges in the same particle. The last several terms used Eq. (\ref{eq:Ldef}) to relate how the product of charges carried by a single particle is determined by the single-charge charge density. If the volumes are moving and expanding with the fluid, and if the currents $\bm{j}_a$ are defined relative to the fluid, one can use the divergence theorem to rewrite Eq. (\ref{eq:QQQlong}) in differential form with $d/dt$ replaced by $D_t$,
\begin{eqnarray}
\label{eq:Dtlhs}
D_t \mathcal{C}^{\rm(tot)}_{abc}(\bm{r}_1,\bm{r}_2,\bm{r}_3,t)
&=&-\nabla_1\cdot\langle\bm{j}_a(\bm{r}_1,t)\delta\rho_b(\bm{r}_2,t)\delta\rho_c(\bm{r}_3,t)\rangle'\\
\nonumber
&&-\nabla_2\cdot\langle\bm{j}_b(\bm{r}_2,t)\delta\rho_a(\bm{r}_1,t)\delta\rho_c(\bm{r}_3,t)\rangle'\\
\nonumber
&&-\nabla_3\cdot\langle\bm{j}_c(\bm{r}_3,t)\delta\rho_a(\bm{r}_1,t)\delta\rho_b(\bm{r}_2,t)\rangle'\\
\nonumber
&&-\nabla_{12}\cdot\left[L^{(2)}_{ab,d}(\bm{r}_{12},t)\cdot\langle\bm{j}_d(\bm{r}_{12},t)\delta\rho_c(\bm{r}_3,t)\rangle'\right]\\
\nonumber
&&-\nabla_{13}\cdot\left[L^{(2)}_{ac,d}(\bm{r}_{13},t)\cdot\langle\bm{j}_d(\bm{r}_{13},t)\delta\rho_b(\bm{r}_2,t)\rangle'\right]\\
\nonumber
&&-\nabla_{23}\cdot\left[L^{(2)}_{bc,d}(\bm{r}_{13},t)\cdot\langle\bm{j}_d(\bm{r}_{23},t)\delta\rho_a(\bm{r}_1,t)\rangle'\right]\\
\nonumber
&&-\nabla_{1}\cdot\left[L^{(2)}_{bc,d}(\bm{r}_{13},t)\cdot\langle\delta\rho_d(\bm{r}_{23},t)\bm{j}_a(\bm{r}_1,t)\rangle'\right]\\
\nonumber
&&-\nabla_{2}\cdot\left[L^{(2)}_{ac,d}(\bm{r}_{13},t)\cdot\langle\delta\rho_d(\bm{r}_{13},t)\bm{j}_b(\bm{r}_2,t)\rangle'\right]\\
\nonumber
&&-\nabla_{3}\cdot\left[L^{(2)}_{ab,d}(\bm{r}_{12},t)\cdot\langle\delta\rho_d(\bm{r}_{12},t)\bm{j}_c(\bm{r}_3,t)\rangle'\right].
\end{eqnarray}

Putting all these terms together gives the result for applying $D_t$ to the l.h.s. of Eq. (\ref{eq:3pointparts}),
\begin{eqnarray}
\label{eq:DtCabcfinal}
D_t\mathcal{C}^{\rm(tot)}_{abc}(\bm{r}_1,\bm{r}_2,\bm{r}_3,t)&=&
-\nabla_1\cdot\langle\bm{j}_a(\bm{r}_1)\delta\rho_b(\bm{r}_2)\delta\rho_c(\bm{r}_3)\rangle'
-\nabla_2\cdot\langle\delta\rho_a(\bm{r}_1)\bm{j}_b(\bm{r}_2)\delta\rho_c(\bm{r}_3)\rangle'\\
\nonumber
&&-\nabla_3\cdot\langle\delta\rho_a(\bm{r}_1)\delta\rho_b(\bm{r}_2)\bm{j}_c(\bm{r}_3)\rangle'\\
\nonumber
&&-\nabla_{12}\cdot\left\{L^{(2)}_{ab,d}(\bm{r}_{12},t)\langle \bm{j}_d(\bm{r}_{12},t)\delta\rho_c(\bm{r}_3)\rangle\right\}
-\nabla_{13}\cdot\left\{L^{(2)}_{ac,d}(\bm{r}_{13},t)\langle \bm{j}_d(\bm{r}_{13},t)\delta\rho_b(\bm{r}_2)\rangle\right\}\\
\nonumber
&&-\nabla_{23}\cdot\left\{L^{(2)}_{bc,d}(\bm{r}_{23},t)\langle \bm{j}_d(\bm{r}_{23},t)\delta\rho_a(\bm{r}_1)\rangle\right\}\\
\nonumber
&&-\nabla_3\cdot\left\{L^{(2)}_{ab,d}(\bm{r}_{12},t)\langle \bm{j}_d(\bm{r}_{12},t)\delta\rho_c(\bm{r}_3)\rangle\right\}
-\nabla_2\cdot\left\{L^{(2)}_{ac,d}(\bm{r}_{12},t)\langle \bm{j}_d(\bm{r}_{13},t)\delta\rho_b(\bm{r}_2)\rangle\right\}\\
\nonumber
&&-\nabla_1\cdot\left\{L^{(2)}_{bc,d}(\bm{r}_{23},t)\langle \bm{j}_d(\bm{r}_{23},t)\delta\rho_a(\bm{r}_1)\rangle\right\}.
\end{eqnarray} 
Next, one applies $D_t$ to the r.h.s. of Eq. (\ref{eq:3pointparts}). First, a sample term is considered where two of the charges are carried by the same particle,
\begin{eqnarray}
\label{eq:Dtrhs}
D_tC^{(2;1)}_{ab;c}(\bm{r}_{12},\bm{r}_3,t)&=&D_t\left[L^{(2)}_{ab,d}(\bm{r}_{12},t)C^{(1;1)}_{d;c}(\bm{r}_{12},\bm{r}_3,t)\right]\\
\nonumber
&=&L^{(2)}_{ab,d}(\bm{r}_{12},t)\left[D_t\chi^{(2)}_{dc}(\bm{r}_3,t)\right]\delta(\bm{r}_{12}-\bm{r}_3)
-L^{(2)}_{ab,d}(\bm{r}_{12},t)\nabla_{12}\cdot\langle \bm{j}_d(\bm{r}_{12},t)\delta\rho_c(\bm{r}_3)\rangle'\\
\nonumber
&&+[(\partial_t+\bm{v}\cdot\nabla_{12})L^{(2)}_{ab}(\bm{r}_{12},t)]C^{(1;1)}_{d;c}(\bm{r}_{12},\bm{r}_3,t)\\
\nonumber
&=&L^{(2)}_{ab,d}(\bm{r}_{12},t)\left[D_t\chi^{(2)}_{dc}(\bm{r}_{12},t)\right]\delta(\bm{r}_{12}-\bm{r}_3)
-\nabla_{12}\cdot\left[L^{(2)}_{ab,d}(\bm{r}_{12},t)\langle \bm{j}_d(\bm{r}_{12},t)\delta\rho_c(\bm{r}_3)\rangle'\right]\\
\nonumber
&&-\nabla_3\cdot \left[L^{(2)}_{ab,d}(\bm{r}_{12},t)\langle \delta\rho_d(\bm{r}_{12},t)\bm{j}_c(\bm{r}_3)\rangle'\right]
+[d_tL^{(2)}_{ab,d}(\bm{r}_{12},t)]C^{(1;1)}_{d;c}(\bm{r}_{12},\bm{r}_3,t).
\end{eqnarray}
Here, the definition of  $d_t$ includes the statistical average to its right, $\langle\delta\rho(\bm{r},t)X\rangle$,
\begin{eqnarray}
\label{eq:dtdef}
[d_tL^{(2)}_{ab,d}(\bm{r},t)]\langle \delta\rho_d(\bm{r},t)X\rangle&&\\
\nonumber
&&\hspace{-60pt}=
\left[\left(\partial_t+\bm{v}(\bm{r},t)\cdot\nabla
+\frac{\langle \bm{j}_d(\bm{r},t)X\rangle}{\langle \delta\rho_d(\bm{r},t)X\rangle}\cdot\nabla\right)
L^{(2)}_{ab,d}(\bm{r},t)\right]\langle \delta\rho_d(\bm{r},t)X\rangle.
\end{eqnarray}
Here, $X$ could refer to any operator away from the position $\bm{r}$. The quantitiy $\langle{\bm j}(\bm{r},t)_dX\rangle$ is reexpressed as a ratio over $\langle\delta\rho_d(\bm{r},t)X\rangle$ multiplied the same quantity. This is motivated so that one can see that $d_t$ is effectively the co-moving derivative, but co-moving in the frame of the current, which is not necessarily the same as the frame of the fluid. Thus, if $\delta\rho$ is represented by Monte Carlo sampling, the derivative $d_t$ would refer to the rate of change according to an observer moving with the sampling particles. 

Comparing Eq. (\ref{eq:Dtlhs}) to Eq. (\ref{eq:Dtrhs}) one can see that many of the terms cancel. The resulting equation expresses the evolution of $C^{(1;1;1)}$,
\begin{eqnarray}
\label{eq:DtC111}
D_t C_{a;b;c}^{(1;1;1)}(\bm{r}_1,\bm{r}_2,\bm{r}_3,t)&=&
-\nabla_1\cdot\langle\bm{j}_a(\bm{r}_1,t)\delta\rho_b(\bm{r}_2,t)\delta\rho_c(\bm{r}_3,t)\rangle'
-\nabla_2\cdot\langle\bm{j}_b(\bm{r}_2,t)\delta\rho_a(\bm{r}_1,t)\delta\rho_c(\bm{r}_3,t)\rangle'\\
\nonumber
&&-\nabla_3\cdot\langle\bm{j}_c(\bm{r}_3,t)\delta\rho_a(\bm{r}_1,t)\delta\rho_b(\bm{r}_2,t)\rangle'\\
\nonumber
&&+S^{(3)}_{abc}(\bm{r}_{123},t)\delta(\bm{r}_1-\bm{r}_2)\delta(\bm{r}_{12}-\bm{r}_3)
+S^{(2;1)}_{ab;c}(\bm{r}_{12},\bm{r}_3,t)\delta(\bm{r}_1-\bm{r}_2)\\
\nonumber
&&+S^{(2;1)}_{ac;b}(\bm{r}_{13},\bm{r}_2,t)\delta(\bm{r}_1-\bm{r}_3)
+S^{(2;1)}_{ab;c}(\bm{r}_{23},\bm{r}_1,t)\delta(\bm{r}_2-\bm{r}_3).\\
\nonumber
S^{(3)}_{abc}(\bm{r},t)&=&-D_t\chi^{(3)}_{abc}(\bm{r},t)-L^{(2)}_{ab,d}S^{(2)}_{cd}(\bm{r},t)\\
\nonumber
&&-L^{(2)}_{ac,d}(\bm{r},t)S^{(2)}_{bd}(\bm{r},t)-L^{(2)}_{bc,d}(\bm{r},t)S^{(2)}_{ad}(\bm{r},t)(\bm{r},t).\\
\nonumber
S^{(2;1)}_{ab;c}(\bm{r},\bm{r}',t)&=&-[d_tL^{(2)}_{ab,d}(\bm{r},t)]C^{(1;1)}_{d;c}(\bm{r},\bm{r}',t).
\end{eqnarray}
The first three terms in Eq. (\ref{eq:DtC111}) describe how the correlations evolve when all three coordinates differ. If the current is diffusive, $\bm{j}_a=-\mathcal{D}_{ab}\nabla\delta\rho_b$, the correlations spread with time. The remaining terms represent source terms for $C^{(1;1;1)}$. In the absence of the source terms $C^{(1;1;1)}$ would integrate to a constant. The last four terms describe the sourcing of $C^{(1;1;1)}$ for instances when at least two of the coordinates are equal. The term proportional to $D_t\chi^{(3)}$ was expected because $\chi^{(3)}$ describes the correlation when all three charges are on the same particle. The three terms proportional to $L^{(2)}S^{(2)}$ describe how some of the correlation of $\delta Q_a\delta Q_b\delta Q_b$ is absorbed by the change of the two-point function, i.e. two of the charges are on one particle and the third on a second particle. The sources $S^{(2;1)}$ describe how the three point function can be seeded with two points on the same particle, and one on a separated particle. The factor $d_tL^{(2)}$ describes how the correlation of two charges carried by one particle split onto two particles if $L^{(2)}=\chi^{(3)}[\chi^{(2)}]^{-1}$ would change with time. In the next section, a graphical scheme is presented, which provides some visual delineation of the various terms above, while providing the means to write down the corresponding terms for four-point or $n-$point correlations.

\section{Graphical representations and higher-order correlations}
\label{sec:npoint}

In the previous section, equations of motion were found for three-point correlation functions in Eq. (\ref{eq:DtC111}). Combined with the expressions for two-point functions in Eq. (\ref{eq:evolveC2}), and using Eq. (\ref{eq:Ldef}), one can find all correlations of order $\delta \rho^3$. The evolution of the two and three-point functions, described in Eq.s (\ref{eq:evolveC2}) and  (\ref{eq:DtC111}), can also be expressed graphically. The elements of the graphs are lines connected by vertices, with the vertices having either zero or one incoming lines and $n$ outgoing lines. The lines will connect space time points $\bm{r}_1,t_1$ and $\bm{r}_2,t_2$ and are Green's functions describing how charge a charge $\delta Q_a$, placed at $\bm{r}_1,t_1$ would affect the density, $\delta\rho_b$, at a point $\bm{r}_2,t_2$, where $t_2>t_1$.
\begin{eqnarray}
\langle\delta\rho_b(\bm{r}_2,t_2)\rangle&=&G_{ab}(\bm{r}_1,t_1,\bm{r}_2,t_2)\delta Q_a.
\end{eqnarray}
The Green function is normalized,
\begin{eqnarray}
\int d^3r_2 G_{ab}(\bm{r}_1,t_1,\bm{r}_2,t_2)&=&\delta_{ab},
\end{eqnarray}
and obeys the boundary condition at $t_1=t_2$, 
\begin{eqnarray}
G_{ab}(\bm{r}_1,t_1,\bm{r}_2,t_2=t_1)&=&\delta(\bm{r}_1-\bm{r}_2)\delta_{ab}.
\end{eqnarray}
For a diffusive equation, $\bm{j}_a=-\mathcal{D}_{ab}\nabla\delta\rho_b$, the Green's function can be calculated by solving the differential equation,
\begin{eqnarray}
D_2G_{ab}(\bm{r}_1,t_1,\bm{r}_2,t_2)&=&-\mathcal{D}_{bc}(\bm{r}_2,t_2)\nabla_2^2G_{ac}(\bm{r}_1,t_1,\bm{r}_2,t_2),\\
\nonumber
D_2&=&\frac{\partial}{\partial t_2}+(\nabla_2\cdot\bm{v}(\bm{r}_2,t_2))+\bm{v}(\bm{r}_2,t_2)\cdot\nabla_2.
\end{eqnarray}
\begin{figure}
\centerline{\includegraphics[width=0.35\textwidth]{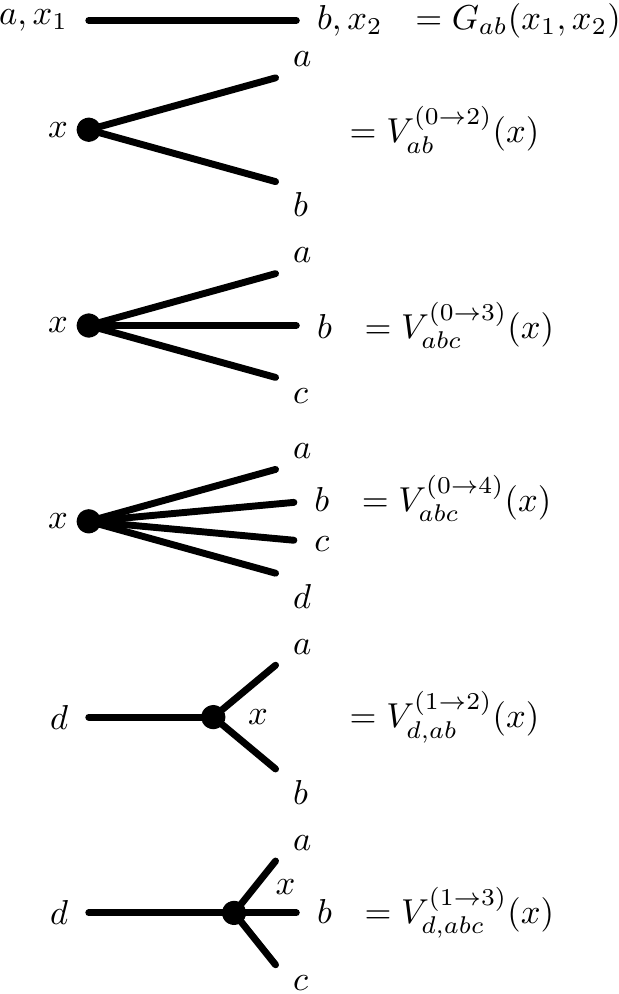}}
\caption{\label{fig:gvdef}
Elements of the graphical representation are defined in Eq. (\ref{eq:gvdef}), with $x$ referring to the space-time point $\bm{r},t$.}
\end{figure}
For any realistic dynamic system, it is unlikely $G$ can be found analytically. The choices are either to solve the differential equation numerically on a three-dimensional mesh, or to sample the diffusive spread as a random walk. The sources of the Green's function for two- and three-point functions are listed in Eq.s (\ref{eq:evolveC2}) and (\ref{eq:DtC111}) respectively. Sources for the Green's functions can be represented diagramatically, with vertices representing sources at points in space time, and lines between the vertices representing Green's functions. These graphical elements are illustrated in Fig. \ref{fig:gvdef} and the vertices are defined below,
\begin{eqnarray}
\label{eq:gvdef}
V^{(0\rightarrow 2)}_{ab}(\bm{r},t)&=&-D_t\chi^{(2)}_{ab}(\bm{r},t),\\
\nonumber
V^{(0\rightarrow 3)}_{abc}(\bm{r},t)&=&-D_t\chi^{(3)}_{abc}(\bm{r},t)-L^{(2)}_{ab,d}V^{(0\rightarrow 2)}_{cd}(\bm{r},t)\\
\nonumber
&&-L^{(2)}_{ac,d}(\bm{r},t)V^{(0\rightarrow 2)}_{bd}(\bm{r},t)-L^{(2)}_{bc,d}(\bm{r},t)V^{(0\rightarrow 2)}_{ad}(\bm{r},t)(\bm{r},t),\\
\nonumber
V^{(0\rightarrow 4)}_{abcd}&=&-D_t\chi^{(4)}_{abcd}\\
\nonumber
&&-L^{(2)}_{ab,e'}(\bm{r},t)V^{(0\rightarrow 3)}_{cd,e'}(\bm{r},t)
-L^{(2)}_{ac,e'}(\bm{r},t)V^{(0\rightarrow 3)}_{bd,e'}(\bm{r},t)
-L^{(2)}_{ad,e'}(\bm{r},t)V^{(0\rightarrow 3)}_{bc,e'}(\bm{r},t)\\
\nonumber
&&-L^{(2)}_{bc,e'}(\bm{r},t)V^{(0\rightarrow 3)}_{ad,e'}(\bm{r},t)
-L^{(2)}_{bd,e'}(\bm{r},t)V^{(0\rightarrow 3)}_{ac,e'}(\bm{r},t)
-L^{(2)}_{cd,e'}(\bm{r},t)V^{(0\rightarrow 3)}_{ab,e'}(\bm{r},t)\\
\nonumber
&&-L^{(2)}_{ab,e'}(\bm{r},t)L^{(2)}_{cd,f'}(\bm{r},t)V^{(0\rightarrow 2)}_{e'f'}(\bm{r},t)
-L^{(2)}_{ac,e'}(\bm{r},t)L^{(2)}_{bd,f'}(\bm{r},t)V^{(0\rightarrow 2)}_{e'f'}(\bm{r},t)\\
\nonumber
&&
-L^{(2)}_{bc,e'}(\bm{r},t)L^{(2)}_{ad,f'}(\bm{r},t)V^{(0\rightarrow 2)}_{e'f'}(\bm{r},t)\\
\nonumber
&&-L^{(3)}_{abc,e'}(\bm{r},t)V^{(0\rightarrow 2)}_{e'd}(\bm{r},t)
-L^{(3)}_{abd,e'}(\bm{r},t)V^{(0\rightarrow 2)}_{e'c}(\bm{r},t)\\
\nonumber
&&-L^{(3)}_{acd,e'}(\bm{r},t)V^{(0\rightarrow 2)}_{e'b}(\bm{r},t)
-L^{(3)}_{bcd,e'}(\bm{r},t)V^{(0\rightarrow 2)}_{e'a}(\bm{r},t),\\
\nonumber
V^{(1\rightarrow 2)}_{d,ab}(\bm{r},t)&=&-d_tL^{(2)}_{ab,d}(\bm{r},t),\\
\nonumber
V^{(1\rightarrow 3)}_{d,abc}(\bm{r},t)&=&-d_tL^{(3)}_{abc,d}(\bm{r},t)-L^{(2)}_{bc,e}(\bm{r},t)V^{(1\rightarrow 2)}_{d,ae}(\bm{r},t)\\
\nonumber
&&-L^{(2)}_{ac,e}(\bm{r},t)V^{(1\rightarrow 2)}_{d,be}(\bm{r},t)
-L^{(2)}_{ab,e}(\bm{r},t)V^{(1\rightarrow 2)}_{d,ce}(\bm{r},t).
\end{eqnarray}

\begin{figure}
\centerline{\includegraphics[width=0.7\textwidth]{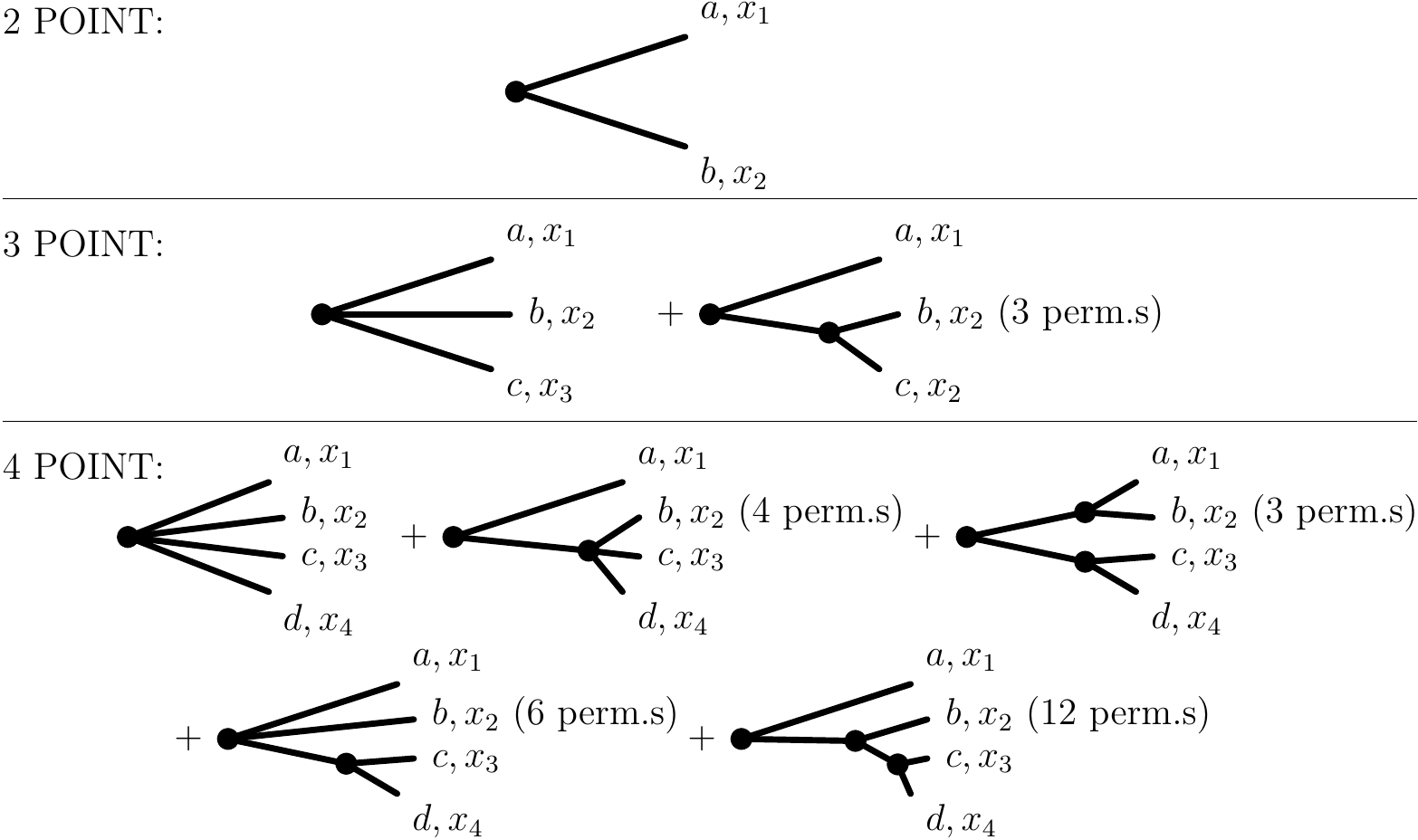}}
\caption{\label{fig:order4}
Diagrams for calculating two-, three- and four-point functions. For topologically identical diagrams which differ by permutations of the final-state labels, the net number of permutations is listed rather than repeating the similar diagrams. Each vertex is assigned a space-time point, over which is integrated.}
\end{figure}

Figure \ref{fig:order4} shows the diagrams for calculating two-point, three-point and four-point functions. Many of the diagrams are topologically identical and are related by permuting the final-state labels. In those cases the similar diagrams are noted by the number of permutations for that topology.
As an example, the contribution to the three-point diagram from the second three-point diagram in Fig. \ref{fig:order4} is the integral
\begin{eqnarray}
C^{(1;1;1)}_{a;b;c}(x_1,x_2,x_2)&=&
\cdots+\int d^4y_1d^4y_2 V^{(0\rightarrow 2)}_{a'd'}(y_1)G_{a'a}(y_1,x_1)G_{d'd}(y_1,y_2)\\
\nonumber
&&V^{(1\rightarrow 2)}_{d,b'c'}(y_2)G_{b'b}(y_2,x_2)G_{c'c}(y_2,x_3).
\end{eqnarray}
Each vertex in the diagram is assigned a space-time point, in this case $y_1$ and $y_2$. Integrations are performed over those coordinates. Each internal line is assigned two charge indices which then determine the charge indices for the vertices. All diagrams begin with a vertex $V^{(0\rightarrow n)}$, and end with open Green's functions denoted by the desired measurement. 

\section{Relation to Charge Fluctuations}

Within some large volume $V$, charge fluctuations are defined
\begin{eqnarray}
F^{(2)}_{ab}&\equiv&\frac{1}{V}\langle \delta Q_a\delta Q_b\rangle\\
\nonumber
F^{(3)}_{abc}&\equiv&\frac{1}{V}\langle \delta Q_a\delta Q_b\delta Q_c\rangle\\
\nonumber
F^{(4)}_{abcd}&\equiv&\frac{1}{V}\langle \delta Q_aQ_b\delta Q_c\delta Q_d\rangle-\frac{1}{V}\langle \delta Q_a\delta Q_b\rangle\langle \delta Q_c\delta Q_d\rangle
-\frac{1}{V}\langle\delta Q_a\delta Q_c\rangle\langle\delta Q_b\delta Q_d\rangle\\
\nonumber
&&-\frac{1}{V}\langle\delta Q_a\delta Q_d\rangle\langle\delta Q_b\delta Q_c\rangle.
\end{eqnarray}
Each charge $Q_a$ can expressed as an integral over the charge density $\delta\rho_a$. For the order $Q^n$ fluctuation, one obtains contributions from the two-point, three-point, up to $n-$point functions. The contribution from the $n-$point function is simply the integral over all the external coordinates in the diagrams from Fig. \ref{fig:order4}. The contributions from the $(n-1)-$point functions with final-state charge indices $a$ and $b$ can be found by attaching an operator $L^{(2)}_{ab,a'}(x)$ to any external Green's function $G_{d'a'}(y,x)$ where $x$ is a final-state coordinate and $a'$ denotes the measured charge. Thus, each 3-point diagram from Fig. \ref{fig:order4} contributes to $F^{(4)}$. The contributions to $F^{(4)}$ from two-point functions come from either attaching $L^{(2)}$ to both of the external lines, or by attaching $L^{(3)}$ to either external line. Finally, $F^{(4)}$ has a contribution from all four charges being on the same particle, which would be represented by $\chi^{(4)}$.

Experimentally, the contributions to $F^{(4)}$ from four-point functions come from summing over all combinations of four final-state particles, never using the same particle twice in the same term. The contribution to $F^{(4)}$ from three-point functions would be found by summing over all sets of three final-state particles then requiring one particle to provide two powers of the charge. The contributions from two-point functions describes the case where the sum extends over all pairs, with each particle contributing an order $Q^2$ contribution or for one particle to provide an order $Q$ and the second providing an order $Q^3$ contribution. Finally, summing over the particles individually, one would add the contribution of $Q_aQ_bQ_cQ_d$ for that particle. Aside from the contribution to $F^{(n)}$ from the $n-$point function, all other contributions are determined by correlations of fewer coordinates, and thus do not represent additional information beyond what would have been gathered by $(n-1)-$point functions.

\section{Algorithms}
\label{sec:algorithms}

In principle, one could solve the differential equations for the correlation functions. The differential equation would involve solving for all points on a grid with three spatial dimensions and one time dimension. If the three-dimensional space-time grid was represented by $N\times N\times N$ grid points calculated for $N_t$ values of the time, an $n$-point correlation function would involve of the order $N^{3n}N_t$ grid points. This would be likely be prohibitively expensive.

Another possibility for calculating $n$-point correlations would be to solve $n$ separate one-point diffusion equations on $n$ meshes. For each source point, one would increment the correlations on each of the meshes. For each source point, $S^{(n)}_{ab\cdots c}d^4x$, one could increment the charges on the corresponding mesh points by amounts $\delta Q_a,\cdots \delta Q_c$, such that the product of the charges reproduced $S^{(n)}_{ab\cdots c}d^4x$, but so that the incremented charge on any of the individual mesh had a random sign. At the final time, one would construct the $n-$point correlation function by using the charges from each of the $n$ meshes. Unless the contributions came from the same source point, they would, on average, cancel. Unfortunately, the cancellation would require repeating the procedure many times to combat combinatoric noise. This would be especially true for $n>2$. For $n>2$ it would be more efficient to evolve the contribution from a single source-point, then construct the correlation. One would perform a Monte Carlo sampling over the many source points using $|S^{(n)}_{ab\cdots c}d^4x|$ as the probability to choose the sampling points, then use $S^{(n)}_{ab\cdots c}/|S^{(n)}_{ab\cdots c}|$ as a weight to increment the correlations functions.

A second approach, built on the assumption that the dynamics is diffusive, is to represent the correlations with clusters of sample particles undergoing random walks. For $n-$point correlations, the clusters would involve $n$ charges. Sample charges move with some velocity $\bm{v}$ and then have their directions reoriented randomly. The probability that a particle is thus scattered during a time interval $dt$ is $dt/\tau$, with $\tau=6\mathcal{D}/v^2$. In the limit $v\rightarrow\infty$ the random walk approaches the diffusion equation. By setting $v$ to the speed of light, it is causal and approaches the diffusion equation after several scatterings. If the diffusivity, $\mathcal{D}_{ab}$, is not diagonal, a more sophisticated representation would need to be invoked, and is described below. Each cluster of particles would be created via a Monte Carlo procedure weighted by the source function, and would evolve as a random walk to mimic the diffusion equation. Each particle carries a unit charge, and each group of particles would carry a weight, which could be either positive or negative, describing the contribution from the original vertex after accounting for the Monte Carlo weight. When calculating the correlations, only those particles within the same cluster need to be combined, which results in low combinatoric noise. Such an approach was applied in \cite{Pratt:2019pnd,Pratt:2018ebf,Pratt:2017oyf}. In those instances, only two-point functions were considered. Calculations for three-and four-point functions would involve accounting for a larger number of diagrams. One advantage of this approach is that vertices of the form $V^{(1\rightarrow n)}$ would be rather straight forward to calculate. These vertices behave as $d_tL^{(n)}_{a\cdots c,b}(x)$. The derivative $d_t$, defined in Eq. (\ref{eq:dtdef}), is the time-derivative that co-moves with the current, or in this case is co-moving with the sampling particles. Thus, thus the $(1\rightarrow n)$ vertices involve calculating how $L_{a\cdots c,d}$ changes according to an observer moving with the sampling particles. This simplifies sampling the secondary vertices with Monte Carlo.

For either approach, the $n-$point functions must be addressed in order. The two-point evolution can be used to calculate the three-point evolution, and the two- and three-point correlations serve as a basis for the four-point function. For the random-walk representation, one stores the correlated clusters of correlated particles. When evolving a pair of particles to represent the two-point function, one could bifurcate one of the particles carrying charge $d$ into two with charges $a$ and $b$ during a time interval $dt$ with probability $V^{(1\rightarrow 2)}_{d,ab}(\bm{r},t)dt$. One would continue to simulate the non-split trajectory for calculation of the two-point function, and would add the bifurcated trajectory into a list of samplings for the three-point function. Such a three-point trajectory, generated from an initially two-point trajectory, would represent the second diagram for 3-point correlations in Fig. \ref{fig:order4}. This would be added to the purely three-point trajectory described by the first three-point diagram in Fig. \ref{fig:order4}. Similarly, one can calculate four-point functions. Given the lack of combinatoric noise, such a calculation would be tenable and require only modest computational resources. However, the simple random walk approach needs to be altered if the diffusivity is not diagonal, as described below. 

\subsection{Non-diagonal diffusivity matrix}

It is straight-forward to model the diffusive evolution to the $n-$point contribution of the $\delta\rho^n$ correlation, $C^{(1;1;\cdots 1)}_{a;b;\cdots c}(\bm{r}_1,\bm{r}_2,\cdots\bm{r}_n,t)$, for the case where the diffusivity matrix is diagonal, if one is given the source function $S^{(n)}_{ab\cdots c}(\bm{r}_1,\bm{r}_2\cdots\bm{r}_n)$. Because diffusion represents a random walk, one simply creates a set of unit charges, $a\cdots c$ at the points $\bm{r}_1\cdots\bm{r}_n$, and assigns a weight to the group. The weight is given by $S^{(n)}_{ab\cdots c}d^4x/P_{MC}$. The probability $P_{MC}$ accounts for the fact that in each four-volume element $d^4x$, one may choose whether or not to create the sampling charges. If the Monte Carlo probability, $P_{MC}$, is chosen as $|S^{(n)}_{ab\cdots c}d^4x|$, then the weights are $\pm 1$, depending on whether the sources are positive or negative. Each charge is then propagated as a random walk. The sample charges move with velocity $v$ relative to the medium, and with random directions. The charges then reorient randomly according to a lifetime, $\tau_{\rm coll}=6\mathcal{D}/v^2$. I.e., in each time step $dt$ the particle reorients with probability $dt/\tau_{\rm coll}$. 

There are significant advantages to using a random walk representation of the diffusion equation. First, such implementations tend to be simpler to implement than the solving the differential equation on a four-dimensional space-time grid. Second, one can make the evolution causal by setting $v$ to the speed of light. In the limit of $v\rightarrow\infty$ the random walk exactly reproduces the diffusion equation, but by lowering $v$ the method prunes the acausal tail of the correlation function. The difference between causal and acausal treatments matters only for short diffusion times. For long times, the random walk approaches the solution to the diffusion equation as long as the evolutions involve many reorientations for each test charge. For the calculations in \cite{Pratt:2017oyf,Pratt:2018ebf,Pratt:2019pnd}, the number of such reorientations was on the order of a half dozen. Finally, the random walk makes it easy to label the contribution to the correlation from the same source point. Contributions from different source points should cancel, so by only incrementing contributions from the same source point the combinatoric noise is greatly reduced. This becomes increasingly important as one considers correlations of increasing order. 

Treating diffusion as a random walk is more complicated once the diffusivity tensor becomes non-diagonal,
\begin{eqnarray}
\bm{j}_a&=&-\mathcal{D}_{ab}\nabla\rho_b.
\end{eqnarray}
The method will be based on considering sample charges in a basis where $\mathcal{D}$ is diagonal. Here, the eigenvectors of $\mathcal{D}$ are labeled $u^{(i)}$. The source functions and susceptibilities can be expressed in this new basis, and the sample charges are labeled by eigenvectors. If the basis were constant throughout the evolution, the algorithm would then be unchanged from what was described above. If one applies Eq. (\ref{eq:dNh}) to translate a sample charge into particles of a specific species, the unit sample charge $\delta Q_a$ is $u^{(i)}_a$, where $i$ refers to the specific eigenvector representing the sample charge. 

The non-diagonal elements are thus rather straight-forward to accommodate if the eigenvectors of the diffusivity tensor do not change as the sample charge traverses the medium. However, when the eigenvectors transform they must be reformulated in terms of the new eigenvectors. Let's assume the original normalized eigenvectors were $\hat{a}$, $\hat{b}$ and $\hat{c}$. The new eigenvectors will be $\hat{a}'$, $\hat{b}'$ and $\hat{c}'$. Also, one can assume the charge is originally in the state $u=\hat{a}$.
Using completeness,
\begin{eqnarray}
\hat{a}&=&(\hat{a}\cdot{a}')\hat{a}'+(\hat{a}\cdot{b}')\hat{b}'+(\hat{a}\cdot{c}')\hat{c}'.
\end{eqnarray}
For the Monte Carlo treatment one can probabilistically choose which new eigenvector to use along with an adjustment of the weight so that on average the charge is still in $\hat{a}$. Here, $w$ designates the original weight assigned to the group of sampling charges, and $w'$ will be the new weight after the charge has been re-designated in the new basis. One can generate a random number $r$ such that $0<r<1$. Using $r$, the following algorithm should maintain the continuity of the charge,
\begin{eqnarray}
{\rm if~} (0<r<|(\hat{a}\cdot\hat{a}')|/Z <r) &{\rm ~then}& u\rightarrow \hat{a}',~{\rm and~}w'=wZ(\hat{a}\cdot{a}')/|\hat{a}\cdot{a}'|\\
\nonumber
{\rm else ~if~}~(0<r<(\hat{a}\cdot{b}')\hat{a}'/Z <r) &{\rm ~then}& u\rightarrow \hat{b}',~{\rm and~}w'=wZ(\hat{a}\cdot{n}')/|\hat{a}\cdot{b}'|\\
\nonumber
{\rm else~} &{\rm ~then}& u\rightarrow \hat{c}',~{\rm and~}w'=wZ(\hat{a}\cdot{c}')/|\hat{a}\cdot{c}'|,\\
\nonumber
Z&\equiv&(\hat{a}\cdot{a}')\hat{a}'+(\hat{a}\cdot{n}')\hat{a}'+(\hat{a}\cdot{a}')\hat{v}'.
\end{eqnarray}
If one averages over values of $r$, the result averages to $u\rightarrow (\hat{a}\cdot\hat{a}')\hat{a}'+(\hat{a}\cdot\hat{b}')\hat{b}'+(\hat{a}\cdot\hat{c}')\hat{c}'$, which indeed equals $u=\hat{a}$. One need only check whether to reassign the basis with time steps sufficiently small so that the change in the diffusive movement is small during that time step.

\section{Applicability}
\label{sec:applicability}

The approach here was inspired by understanding how three- and four-point correlations measured in heavy-ion collisions could be modeled. In particular, the goal was to understand the role of local charge conservation. During a heavy-ion collision, one changes phases from a quark-gluon plasma to a hadronic gas. At zero baryon chemical potential, calculations of the susceptibilities from lattice gauge theory suggest that for temperatures below 150 MeV the system is reasonably represented as a hadronic gas, whereas for temperatures above 200 MeV quarks are reasonable quasi-particles. For intermediate temperatures, the transition appears smooth. During a central collision of heavy ions at LHC energies or at RHIC energies, the system traverses a range of temperatures from well above 200 MeV to approximately 100 MeV, and undergoes a radical change in chemistry during that time. The changing number of up, down and strange charges, and the combination of such charges into hadrons induces a rich evolution of charge correlations. If such correlations are short range, and if the expansion is not too fast, it seems a reasonable approximation to assume chemical equilibrium, at least until the temperatures fall below 150 MeV, at which point chemical rates fall below the expansion rate. If the chemistry is equilibrated, one would expect the short range, $\lesssim 1$ fm, correlations to match that of an equilibrated gas. However, beyond one Fermi correlations due to local charge conservation persist. The treatment presented here would thus seem a reasonable approximation to reality, and comparing predictions to measurements would provide a stringent test of the assumption of local chemical equilibrium. Indeed, for two-point correlations this approach has matched a range of experimental measurements. 

If chemical equilibrium is not attained the approach can become invalid. For two-point correlations one can assume the local correlation is some function $\chi_{ab}\sim\delta(\bm{r}-\bm{r}')$, and if the function $\chi_{ab}$ can be modeled the approach can still be applied as chemical equilibrium was not an essential approximation. As long as the correlation is local, compared to the size of the system, the same approach, but with a different model for the local correlation, would remain valid. However, the derivations for three- and four-body correlations did rest on the assumptions of chemical equilibrium. Even if one had a model of the local correlations, $\chi^{(3)},\chi^{(4)},\cdots$, that would not be sufficient to understand how a charge $\langle\delta Q_a\rangle$ on a particle would translate into knowing $\delta Q_b\delta Q_c$ on the same particle. Local chemical equilibrium was critical in deriving Eq. (\ref{eq:Ldef}).

Aside from chemical equilibrium, the second assumption is that the correlation is sufficiently local to separate it from the balancing correlation. This should be true in most cases, but would fail for correlations associated with phase transitions. In the critical region correlations fall as power laws, effectively with infinite extent. When inside the coexistence region, bubbles and drops represent macroscopic structures that should not be described by an expansion of $\langle\delta\rho^n\rangle$. For high-energy heavy-ion physics, there remains the possibility that a phase transition exists at finite baryon density, and might be accessible at the lower range of beam energies at RHIC. If the correlations from bulk structure are sufficiently long range, it is possible that short-range correlations, e.g. those from charge conservation, might be superimposed onto a model where a one-body description \cite{Steinheimer:2013xxa,Steinheimer:2013gla,Steinheimer:2012gc,Randrup:2010ax,Heiselberg:AnnPhys,Chomaz:2003dz,Borderie:2001jg,Colonna:2002ti,Guarnera:1996svb,Paech:2005cx,Nahrgang:2011mg}, including some with implementations of noise \cite{Napolitani:2014ima,Gavin:2016hmv,Kapusta:2012sd,Young:2014pka,Pratt:2017lce}. The strategy would be then to first treat the bulk correlations, including the critical correlations and those related to phase separation, using some form of hydrodynamics. Highly local correlations, including their contribution to the susceptibilities, would be ignored for this first pass. The methods presented here could then be applied to account for the remainder of the correlation, i.e. those from short-range correlations and the associated balancing charge. 

Even if one's main motivation for analyzing multi-charge correlations and fluctuations is to search for evidence of phenomena related to phase transitions, it is crucial to estimate the degree to which the short-range correlations and the associated charge balance affect the result. The methods presented here provide a means to calculate that background. If one's goal is to investigate the chemical evolution of a heavy-ion collision through $n-$point correlations in a system where there are only short-range correlations, the methods here make it possible to extend such studies to $n>2$.

\begin{acknowledgments}
This work was supported by the Department of Energy Office of Science through grant number DE-FG02-03ER41259, and benefited from conversations within the Beam Energy Scan Theory (BEST) Topical Collaboration, also supported by the Department of Energy.
\end{acknowledgments}

\end{document}